\documentclass[prd,twocolumn,showpacs,nofootinbib,superscriptaddress]{revtex4}

\usepackage{times}
\usepackage{natbib}
\usepackage{epsfig}
\usepackage{bm}
\usepackage{amsmath}
\usepackage{amssymb}

\newcommand{\NN}{N}            
\newcommand{\CP}{\varphi}
\newcommand{\II}{\mathcal{R}}  
\newcommand{\up}[1]{{\rm #1}}
\newcommand{\beeq}{\begin{equation}}
\newcommand{\eneq}{\end{equation}}
\newcommand{\bear}{\begin{eqnarray}}
\newcommand{\enar}{\end{eqnarray}}
\newcommand{\nnn}{\nonumber \\}

\newcommand{\fnl}{f_\up{NL}}
\newcommand{\gnl}{g_\up{NL}}
\newcommand{\OO}{\mathcal{O}}

\newcommand{\gbar}{\bar g}     
\newcommand{\UU}{u}

\begin{document}

\title{Relativistic effects and primordial non-Gaussianity in the matter density fluctuation}

\author{Jaiyul Yoo}
\email{jyoo@physik.uzh.ch}
\affiliation{Center for Theoretical Astrophysics and Cosmology, 
Institute for Computational Science, University of Z\"urich,
CH-8057 Z\"urich, Switzerland}
\affiliation{Physik-Institut, University of Z\"urich,
CH-8057 Z\"urich, Switzerland}

\author{Jinn-Ouk Gong}
\email{jinn-ouk.gong@apctp.org}
\affiliation{Asia Pacific Center for Theoretical Physics, Pohang, 790-784, Korea}
\affiliation{Department of Physics, Postech, Pohang 790-784, Korea}

\begin{abstract}
We present the third-order analytic solution of the matter density
fluctuation in the proper-time hypersurface of nonrelativistic matter flows
by solving the nonlinear general relativistic equations.
The proper-time hypersurface provides a coordinate system that a local
observer can set up without knowledge beyond its neighborhood, along with
physical connections to the local Newtonian
descriptions in the relativistic context. 
The initial condition of our analytic solution is set up by the 
curvature perturbation in the comoving gauge, clarifying its impact on 
the nonlinear evolution.
We compute the effective non-Gaussian parameters due to the 
nonlinearity in the relativistic equations. With proper coordinate rescaling, 
we show that the equivalence principle is respected 
and the relativistic effect vanishes in the large-scale limit.
\end{abstract}


\preprint{APCTP-Pre2015-023}

\maketitle

Rapid developments in large-scale galaxy surveys over the past decades
have enabled the precision measurements of galaxy clustering, which can be
used to probe the nature of dark energy and the perturbation generation 
mechanism in the early Universe \cite{DETF06}.
In parallel, the recent theoretical development 
(\cite{YOFIZA09,YOO10,BODU11,CHLE11,JESCHI12}; see \cite{YOO14a} for review)
has revealed
that the subtle relativistic effects are present in galaxy clustering,
providing new opportunities to extract additional and critical information
about the gravity on large scales and the initial conditions 
for structure formation.
In particular, the relativistic
formalism has been extended \cite{YOZA14,BEMACL14b,DIDUET14}
to the second-order in perturbations for the computation of higher-order
statistics such as the bispectrum.

One of the critical elements in the relativistic
formalism is galaxy bias, which relates the galaxy number density to the
underlying matter distribution. Beyond the linear order in perturbations,
however, galaxy bias poses a nontrivial problem due to 
the gauge issues in general 
relativity. It was shown \cite{YOO14b} that the proper-time hypersurface
of nonrelativistic matter flows provides a physical description of the local
observer, moving with dark matter and baryons that will collapse to form
galaxies. This physical justification has led us to study the
nonlinear relativistic effects of the matter density fluctuation in the
proper-time hypersurface of nonrelativistic matter flows.

This is timely and interesting, for there has been recent conflict
in literature --- It is argued in \cite{BRHIET14,BRHIWA14}
by performing the second-order relativistic calculations that 
the nonlinear evolution of gravity generates the 
local-type non-Gaussianity, which would in turn exhibit a prominent signature
in galaxy clustering on large scales. 
On the other hands, in \cite{PASCZA13,DAPASC15,DEDOGR15}
the opposite claim is asserted 
that the observable quantities
are {\it not} affected by these nonlinear relativistic effects of gravity,
while  the calculations are in general
based on studying the special case (or the squeezed limit),
where only the linear-order calculations are required. In this 
article, 
we present the third-order relativistic
calculations of the matter density fluctuation
in the proper-time hypersurface, providing the essential tool for computing
the relativistic effects in the higher-order galaxy clustering statistics 
and explicitly resolving the hotly debated issues of the nonlinear relativistic
effects in general relativity.
Throughout the article we use $a,b,c,\cdots$ for the spacetime indices and
$i,j,k,\cdots$ for the spatial ones.

Let us consider a Friedmann-Robertson-Walker universe with
an irrotational pressureless medium of nonrelativistic 
matter, encompassing baryons and dark matter on large scales.
A local observer moving with this nonrelativistic matter flow is described
by its four velocity $\UU^a$, and the energy-momentum tensor in this case is
greatly simplified as
$T_{ab}=\rho_m\UU_a\UU_b$, where $\rho_m$ is the energy density 
of the fluid measured by the local observer. 
As our temporal gauge condition,
we choose the {\it comoving gauge}, where local observers see vanishing
energy flux $T^0{}_i=0$. In this case, the global time coordinate is 
synchronized with the proper-time, and the local observer~$\UU^a$ is aligned
with the geometric normal observer~$n^a$ ($n_i=0$) \cite{YOO14b},
so that the local observer moves along the geodesic $\NN=1$,
where $N$ is the lapse function in the Arnowitt-Deser-Misner (ADM)
formalism~\cite{ADM,MTW}.

Given that the local
observer coincides with the normal observer in our 
temporal gauge condition, the expansion~$\theta$ and the shear~$\sigma_{ij}$
of the flow are related to the extrinsic curvature
tensor~$K_{ij}$ of the 3-hypersurface 
\beeq
K_{ij} \equiv \frac{1}{2N} \left( N_{i:j} + N_{j:i} - \dot{h}_{ij} \right)
=-u_{i;j}~,
\eneq
as
\beeq
\label{eq:extrinsic}
-\theta = K = h^{ij}K_{ij} ~,
\quad \text{and} \quad
-\sigma_{ij} = K_{ij}-{1\over3}h_{ij}K~,
\eneq
where $N_i$ is the ADM shift vector, the dot is the time derivative, and
the colon is the covariant derivative with respect to the projection 
tensor $h_{ab} \equiv g_{ab}+u_au_b$, which is the induced spatial metric of
3-hypersurface.

Moreover, as our spatial gauge condition which matters beyond the linear order
in perturbations, we choose the spatial C-gauge condition \cite{NOHW04}
to have only the diagonal part in the spatial metric
\beeq
h_{ij} \equiv a^2(1+2\CP)\gbar_{ij}~,
\eneq
with curvature perturbation~$\CP$ and the scale factor~$a$.
This results in non-vanishing shift vector $g_{0i}=N_i\equiv-\nabla_i\chi$, 
where $\gbar_{ij}$ is the background 3-metric and the covariant 
derivative~$\nabla_i$ is with respect to~$\gbar_{ij}$.
Notice that 
the spatial B-gauge condition combined with our temporal comoving
gauge condition, the so-called comoving-synchronous gauge, 
yields the simplest metric 
\beeq
h_{ij} \equiv a^2 \left[ (1+2\CP)\gbar_{ij} + 2\nabla_i\nabla_j\gamma 
\right] \quad \text{and} \quad
g_{0i}=0~,
\eneq
at the cost of nonvanishing off-diagonal term~$\gamma$ in the spatial metric.
Our choice of gauge conditions corresponds to a global coordinate system
with the proper-time of a local observer, and this choice leaves
no residual gauge mode \cite{YOO14b}.
Note that we assume no vector or tensor perturbation
present in the initial condition.

With our temporal and spatial gauge conditions, the continuity,
the Raychaudhuri, the ADM energy and momentum constraint equations become
the nonlinear dynamical equations for 
the density perturbation $\delta \equiv \rho/\bar\rho_m-1$ and 
the extrinsic curvature perturbation $\kappa \equiv 3H+K$ as
\bear
\label{eq:delta}
&& \dot\delta-\kappa = \NN^i\nabla_i\delta+\kappa~\delta~,
\\
\label{eq:kappa}
&& \dot\kappa+2H\kappa-4\pi G\bar\rho_m\delta =
\NN^i\nabla_i\kappa+{1\over3}\kappa^2+\sigma^{ab}\sigma_{ab}~,
\\
\label{eq:dR}
&& \delta R
= \sigma^{ab}\sigma_{ab}+4H\kappa-{2\over3}\kappa^2+16\pi G\bar
\rho_m\delta~,\\
\label{eq:sigma}
&& \sigma^j_{i:j} = {2\over3}~\nabla_i\kappa~,
\enar
where $\delta R$ is the perturbation of the intrinsic curvature of the 
3-hypersurface and $H=\dot a/a$ is the Hubble parameter.
To the third order in perturbations, each perturbation component can
be readily computed in terms of metric perturbations \cite{MTW,WALD84}. 

The ADM momentum constraint \eqref{eq:sigma} and
extrinsic curvature \eqref{eq:extrinsic} provide auxiliary
equations for the comoving gauge curvature~$\CP$ and the scalar
shear~$\chi$ to be solved for~$\delta$ and~$\kappa$, and they are explicitly
\begin{widetext}
\begin{align}
\label{eq:perturb}
\kappa+{\Delta\over a^2}\chi & = {1\over a^2}\left[2\CP\Delta\chi(1-2\CP)
-\nabla^i\CP\nabla_i\chi(1-4\CP)+{3\over2}\Delta^{-1}\nabla^i
\left(\nabla_i\chi\Delta\CP+\nabla_j\nabla_i\CP\nabla^j\chi\right)\right]
\nonumber\\
& \quad -{3\over a^2}
\Delta^{-1}\nabla^i\left[2\CP\left(\nabla^j\chi\nabla_j\nabla_i \CP
+\Delta\CP\nabla_i\chi\right)
+{1 \over2}\left(\nabla_i\chi\nabla_j\CP
+3\nabla_j\chi\nabla_i\CP\right)\nabla^j\CP\right]~,
\\
\kappa+3\dot\CP+{\Delta\over a^2}\chi & =
6\CP\dot\CP+{1\over a^2}\left[2\CP\Delta\chi(1-2\CP)
-\nabla_i\chi\nabla^i\CP (1-4\CP)\right]~,
\end{align}
where 
$\Delta^{-1}$ is the inverse Laplacian operator.
Combining these two equations with the
geodesic condition~$N=1$, we obtain
the evolution equation for the curvature perturbation
\begin{equation}
\label{eq:td}
\dot\varphi = 2\CP\dot\CP
-{1\over2a^2}\Delta^{-1}\nabla^i\bigg[
\nabla^j\chi\nabla_j\nabla_i\CP+\Delta\CP\nabla_i\chi
-4\CP\left(\nabla^j\chi\nabla_j\nabla_i \CP
+\Delta\CP\nabla_i\chi\right)-\left(\nabla_i\chi\nabla_j\CP
+3\nabla_j\chi\nabla_i\CP \right)\nabla^j\CP\bigg]~,
\end{equation}
then we arrive at the well-known result that the comoving gauge curvature
is conserved at the linear order in perturbations: 
$\CP^{(1)}\equiv\II^{(1)}(\bm x)$, where $\II$ is the initial condition
and the superscript $(n)$ means $n$-th order in perturbation.
To simplify the time evolution, we 
now assume that the universe is matter-dominated
(In a $\Lambda$CDM universe, the time-dependence of the solution is more
complicated. However, the spatial function in Eq.~\eqref{eq:fullsol11} is
identical, leaving our conclusion on the effective non-Gaussian parameters
unaffected.)
 Equation~\eqref{eq:td} can 
then be analytically integrated at each order in perturbations, 
and up to third order the solution is found as
\bear
\label{eq:pt}
\CP(t,{\bm x}) & =& \II + 2\II\CP^{(2)}
-{2\over5(aH)^2} \bigg\{ {1\over4} \nabla_i\II\nabla^i\II
+ {1\over2} \Delta^{-1}\nabla^i \left[ \Delta\II\nabla_i\II
+ \nabla_j\nabla_i\II\nabla^j \left(\frac{5}{2}H\chi_1^{(2)} \right)
+\nabla_i\left(\frac{5}{2}H\chi_1^{(2)}\right)\Delta\II \right] 
\nnn
&& -2\Delta^{-1}\Big[\left(\nabla^i\nabla^j\II\nabla_i\nabla_j\II
+\Delta\II\Delta\II+2\nabla^i\II\Delta\nabla_i\II\right)\II
+3\nabla_i\nabla_j\II\nabla^i\II\nabla^j\II
+2\Delta\II\nabla^i\II\nabla_i\II\Big]\bigg\}
\nnn
&& 
+{1\over4}\Delta^{-1}\nabla^i\left(\nabla_j\nabla_i\II\Delta^{-1}\nabla^j
{\kappa_2^{(2)}\over H}+\Delta^{-1}\nabla_i{\kappa_2^{(2)}\over H}
\Delta\II\right) - {1\over10(aH)^2}\Delta^{-1}\nabla^i
\left(\nabla_j\nabla_i\CP^{(2)}\nabla^j\II+\nabla_i\II\Delta\CP^{(2)}\right) ~,
~~~~~~~~
\enar
with the second-order scalar shear
$5H\chi_1^{(2)}/2 = \Delta^{-1}(\II\Delta\II)/2 
- 3\Delta^{-2}\nabla_i\nabla_j \left( \II\nabla^i\nabla^j\II \right)$
in Eq.~(\ref{eq:perturb})  and 
$\kappa_2^{(2)}/H = [4(aH)^{-4}/175] \left[ 2\Delta \left(
\nabla^i\II\nabla_i\II \right) + 3\nabla^i \left( \Delta\II\nabla_i\II 
\right) \right]$ in Eq.~(\ref{eq:fullsol22}).
The meaning of the subscript will be explained soon.
Beyond the linear order, the curvature perturbation~$\CP$
grows in time due to the nonlinearity in the evolution equation:
$\CP^{(2)}\propto(aH)^{-2}\propto t^{2/3}$ and 
$\CP^{(3)}\propto(aH)^{-4}\propto t^{4/3}$, 
and it all vanishes to all orders in perturbations on superhorizon scales.

Combining Eqs.~\eqref{eq:delta} and~\eqref{eq:dR},
we write the master
differential equation to be solved for~$\delta$:
\beeq
\label{eq:master}
a^2\left(H\dot\delta + {3\over2}H^2\delta\right)
= {a^2\over4} \left( \delta R-\sigma^{ab}\sigma_{ab}+{2\over3}\kappa^2
+4H\NN^i\nabla_i\delta+4H\kappa\delta \right)~,
\eneq
and its relation to~$\kappa$ is given by the ADM energy constraint equation,
written explicitly as
\begin{align}
\label{eq:ADM}
\frac32H^2\delta+H\kappa+{1\over a^2}\Delta\CP&=
{1\over6}\kappa^2+{1\over 12a^4}\left[(\Delta\chi)^2
-3\nabla_i\nabla_j\chi\nabla^i\nabla^j\chi\right](1-4\CP)
+{1\over a^2}\left(4\varphi\Delta\varphi+\frac32
\nabla^i\varphi\nabla_i\varphi\right) 
\nonumber\\
& \quad +{1\over a^4}\left[\nabla^j\nabla^i\chi
\nabla_j\CP\nabla_i\chi-{1\over3}\nabla^i\CP\nabla_i\chi\Delta\chi
\right]-{3\over a^2}\CP\left(3\nabla^i\CP\nabla_i\CP+4\CP\Delta\CP\right)~.
\end{align}
\end{widetext}
The homogeneous solution of Eq.~\eqref{eq:master}
satisfies $\delta_h\propto~H$, corresponding to the usual decaying mode.
The particular solution that corresponds to the growing mode can be obtained
as
\beeq
\label{eq:sol}
\delta_p = \delta_h\int{ dt\over\delta_h} 
\frac{\text{RHS of \eqref{eq:master}}}{a^2H}~.
\eneq
We can straightforwardly compute RHS and arrange it
as a sum of scale-dependent and time-dependent functions:
\begin{equation}
\up{RHS}\equiv\up{RHS}_1(\bm x)+\up{RHS}_2(t,\bm x)+\up{RHS}_3(t,\bm x)~,
\end{equation}
where the spatial and the time dependencies of RHS$_i$ can be further
separated as RHS$_i(t,\bm x)\equiv X_i(\bm x)/(aH)^{2(i-1)}$
and RHS$_i$ vanishes at $n$-th order in perturbations for $n<i$.

Therefore, the particular solution in Eq.~(\ref{eq:sol}) is the sum of 
individual solutions~$\delta_i$ associated with RHS$_i$ in Eq.~(\ref{eq:sol}),
i.e. $\delta_p=\delta_1+\delta_2+\delta_3$, where 
$\delta_i(t,\bm{x})\equiv D_i(t)X_i(\bm{x})$ and
\begin{equation}
D_i(t) = H \int \frac{dt}{(aH)^{2i}} = \frac{1}{(aH)^{2i}(i+3/2)}~.
\end{equation}
It should be emphasized that the subscript $i$ means 
$i$-th time dependence, {\em not} necessarily $i$-th order in perturbation
which is {\em separately} denoted by the superscript $(i)$.
For instance, RHS$^{(1)}_1=X_1^{(1)}=-\Delta\II$ is a time-independent 
spatial function set by the initial condition, and the linear-order solution is
$\delta_1^{(1)}=-D_1\Delta\II$, where the time-dependence
of $D_1 = 2(aH)^{-2}/5 \propto t^{2/3}$
is identical to the Newtonian linear-order growth factor when 
normalized to unity at some epoch. 

\begin{widetext}
Finally, the full third-order solutions are
\begin{align}
\label{eq:fullsol11}
\delta_1(t,\bm x) & = {2\over5(aH)^2}\left[-\Delta\II
+{3\over2}\nabla^i\II\nabla_i\II+4\II\Delta\II
-3\II\left(3\nabla^i\II\nabla_i\II+4\II\Delta\II\right)\right]
={\kappa_1(t,\bm x)\over H}~,\\
\label{eq:fullsol12}
\delta_2(t,\bm x) & = {2^2\over5^2(aH)^4}
\Bigg\{{1\over7}(\Delta\II)^2\left(5+{8 \over 3}\II \right)
+{2\over7}\nabla^i\nabla^j\II\nabla_i\nabla_j\II(1-4\II)
+\nabla^i\II\Delta\nabla_i\II\left(1-2\II\right)
-{8\over7}\nabla^i\nabla^j\II\nabla_i\II\nabla_j\II 
\nnn
& \qquad\qquad
+{8\over21}\nabla^i\II\nabla_i\II\Delta\II
+\left(\Delta\nabla_i\II\nabla^i+{4\over7}\nabla_i\nabla_j\II\nabla^i\nabla^j
-{4\over21}\Delta\II\Delta\right)\left( 
D_1^{-1}\Delta^{-1}\delta_1^{(2)} + \frac{5}{2}H\Delta\chi_1^{(2)}
\right)
\nnn
& \qquad\qquad
-\left({2\cdot5\over7}\Delta\II
+\Delta\nabla_i\II\Delta^{-1}\nabla^{i}+\nabla_i\II\nabla^{i}
+{2\cdot2\over7}\nabla_i\nabla_j\II\Delta^{-1}\nabla^i\nabla^j\right)
D_1^{-1}\delta_1^{(2)}\Bigg\}~,\\
\label{eq:fullsol13}
\delta_3(t,\bm x) & = -{1\over5\cdot 9(aH)^2}\left[
2\Delta\left(\nabla_i\II\Delta^{-1}\nabla^i{\kappa_2^{(2)}\over H}\right)
+7\nabla^i\left(\Delta^{-1}\nabla_i{\kappa^{(2)}_2\over H}\Delta\II\right)
+7\nabla^i\left(\delta_2^{(2)}\nabla_i\II\right)\right]~.
\end{align}
By using Eq.~(\ref{eq:ADM}), the perturbations to the extrinsic curvature are
\begin{align}
\label{eq:fullsol22}
{\kappa_2(t,\bm x)\over H}&={2^2\over5^2(aH)^4}
\Bigg\{{1\over7}(\Delta\II)^2\left(3+{16 \over 3}\II\right)
+{4\over7}\nabla^i\nabla^j\II\nabla_i\nabla_j\II(1-4\II)
+\nabla^i\II\Delta\nabla_i\II\left(1-2\II \right)
-{16\over7}\nabla^i\nabla^j\II\nabla_i\II\nabla_j \II
\nnn
& \qquad
+{16\over21}\nabla^i\II\nabla_i \II\Delta\II
+\left(\Delta\nabla_i\II\nabla^i+{8\over7}\nabla_i\nabla_j\II\nabla^i\nabla^j
-{8\over21}\Delta\II\Delta\right)\left( 
D_1^{-1}\Delta^{-1}\delta_1^{(2)} + \frac{5}{2}H\Delta\chi_1^{(2)}\right)
\nnn
& \qquad
-\left({2\cdot3\over7}\Delta\II
+\Delta\nabla_i\II\Delta^{-1}\nabla^i+\nabla_i\II\nabla^i
+{2\cdot4\over7}\nabla_i\nabla_j\II\Delta^{-1}\nabla^i\nabla^j\right)
D_1^{-1}\delta_1^{(2)}\Bigg\}~,\\
\label{eq:fullsol23}
{\kappa_3(t,\bm x)\over H} & = -{1\over5\cdot 3(aH)^2}\left[
2\Delta\left(\nabla_i\II\Delta^{-1}\nabla^i{\kappa_2^{(2)}\over H}\right)
+\nabla^i\left(\Delta^{-1}\nabla_i{\kappa^{(2)}_2\over H}\Delta\II\right)
+\nabla^i\left(\delta_2^{(2)}\nabla_i\II\right)\right]~.
\end{align}
\end{widetext}
These solutions constitute the full third-order relativistic dynamics in the
proper-time hypersurface of nonrelativistic matter flows. It is well-known
that $\delta_i^{(i)}$ and $\kappa_i^{(i)}$ are identical to the Newtonian
solutions with the standard kernels $F_i$ and $G_i$ in Fourier space.
The remainder $\delta_1^{(2,3)}$ and $\delta_2^{(3)}$ of the solution,
and similarly for $\kappa$,
represent the relativistic corrections. The second-order relativistic
correction $\delta_1^{(2)}$ has been derived in literature 
\cite{VEMA09,BAMAET10,HWNOGO12,UGWA14,BRHIET14,BRHIWA14},
but it is the {\it first time} that the full third-order solution 
in Eqs.~(\ref{eq:fullsol11})$-$(\ref{eq:fullsol23})
is presented.
Furthermore, our solution differs from \cite{NOHW08,JEGOET11} 
in the relativistic corrections $\delta_1^{(2,3)}$ and $\delta_2^{(3)}$,
clarifying the direct connection to the
initial condition set by the curvature potential~$\II$. In particular,
the nonlinear relativistic effects in~$\delta_1^{(2,3)}\propto(aH)^{-2}$ are
at the heart of the recent debate in literature, and we further elaborate
as follows.

In the presence of primordial non-Gaussianities, 
there exists a nontrivial coupling
between long and short wavelength
modes. For example, the local-type non-Gaussianity
in the initial condition is often phrased as,
up to cubic order,
\begin{equation}
\zeta({\bm x}) =\zeta_G(\bm x)+{3\over5}\fnl\zeta_G^2(\bm x)
+{9\over25}\gnl\zeta_G^3(\bm x)~,
\end{equation}
where $\zeta_G$ is a linear-order Gaussian random field and
$e^{2\zeta}\equiv1+2\II$ in our notation convention, i.e. 
$\II=\zeta+\zeta^2+2\zeta^3/3$.
Separating the Gaussian curvature
perturbation into long and short wavelength modes
$\zeta_G = \zeta_l+\zeta_s$, the
curvature perturbation $\zeta_\up{short}$ on small scale 
 including the non-Gaussian contributions is expressed as
\begin{equation}
\label{eq:NGshort}
\zeta_\up{short}
= \zeta_s \left( 1+\frac65\fnl\zeta_l+{27\over25}\gnl
\zeta_l^2 \right) + \OO\left(\zeta_s^2\right)~,
\end{equation}
where the long-short mode coupling is explicit. A similar separation of 
long and short 
wavelength
modes can be performed for the matter density fluctuation
in Eq.~(\ref{eq:fullsol11}),
but to simplify the calculation we assume that a long-
wavelength
mode of interest
is larger than the horizon scale, neglecting its gradient:
\begin{equation}
\label{eq:nonG}
\delta_{1,s}(t,{\bm x}) = \left(1-2\zeta_l+2\zeta_l^2\right) \delta_G 
+ \OO\left( \nabla^i\zeta_s\nabla_i\zeta_s \right)~,
\end{equation}
where $\delta_G \equiv -D_1\Delta\zeta_s$. 
Comparing Eqs.~\eqref{eq:NGshort} with~\eqref{eq:nonG}
shows that 
even in the absence of the primordial non-Gaussianity
$\fnl = \gnl = 0$, the nonlinear evolution
of gravity in general relativity effectively 
generates the non-Gaussianity in the matter fluctuation
$\Delta\fnl = -5/3$ and $\Delta\gnl = 50/27$.
Operationally, the long-short coupling in Eq.~(\ref{eq:nonG})
originates from the relativistic effects like $\II\Delta\II$ and 
$\II^2\Delta\II$ in Eq.~(\ref{eq:fullsol11}) that are inherently present
due  to the nonlinearity of the relativistic constraint equation, even when
the initial condition~$\II$ is Gaussian, and this calculation is the
core argument that supports the nonlinear generation of non-Gaussian signatures
in general relativity  \cite{VEMA09,BAMARI04,BRHIET14,BRHIWA14,BABEET15}.

However, 
the situation is puzzling, for such effects of nonlinear gravity
persist in the superhorizon limit, affecting the small-scale dynamics.
This is in conflict with the equivalence principle --- While 
long-mode fluctuations
{\it do} affect the small-scale dynamics, their impact progressively 
decreases, {\it vanishing} in the large-scale limit, contrary to 
Eq.~\eqref{eq:nonG}, where the small-scale dynamics is affected by the
super-horizon wavelength mode of gravity.
This situation is reminiscent of the consistency relation 
\cite{MALDA03,CRZA04,PASCZA13} in single-field 
inflationary scenarios, in which the bispectrum of~$\zeta$ in the 
squeezed limit is proportional to the spectral index 
of the power spectrum of $\zeta$,
but is in fact
identically vanishing with unobservable rescaling of spatial coordinates.
We will show that the unphysical character of Eq.~\eqref{eq:nonG} is 
removed with {\it constant rescaling} of spatial coordinates and
hence it bears no physical significance.

Noting that~$\zeta_l= \zeta_l({\bm x})$ 
is a time-independent spatial function that varies
negligibly within our horizon, we consider a spatial coordinate rescaling
$d\tilde x^i \equiv e^{\zeta_l}dx^i$, leaving the metric at early times
in the 
{\em same} gauge condition
\begin{equation}
ds^2=-dt^2+a^2e^{2\zeta}dx^2=-dt^2+a^2e^{2\zeta_s(x)}d\tilde x^2~,
\end{equation}
where $g_{0i}$ vanishes at $t\rightarrow0$.
We thus identify the curvature perturbation $\tilde\zeta(\tilde{\bm x})$ 
in the rescaled coordinate as $\tilde\zeta(\tilde{\bm x}) = \zeta_s({\bm x})$.
With the chain rule, we can find
\begin{equation}
\Delta\zeta({\bm x}) = \gbar^{ij}\nabla_i\nabla_j\zeta({\bm x})
 = e^{2\zeta_l}\tilde\Delta\zeta({\bm x})~,
\end{equation}
where the Jacobian factor $e^{2\zeta_l}$ is present 
in addition to the rescaled Laplacian operator $\tilde\Delta$.
Since the matter density fluctuation is a scalar, it remains unchanged
under the spatial coordinate transformation
$\delta(t,{\bm x})=\tilde\delta(t,\tilde{\bm x})$. 
Plugging this into Eq.~\eqref{eq:nonG}, we find
\begin{equation}
\delta_{1,s}(t,{\bm x})
= -D_1\tilde\Delta\tilde\zeta(\tilde{\bm x}
) + \OO \left( \tilde\nabla^i\tilde\zeta
\tilde\nabla_i\tilde\zeta \right)
 \approx 
\delta_G~,
\end{equation}
so that $\delta_{1,s}$
is now explicitly devoid of any correlation to a long-
wavelength
mode fluctuation 
beyond our horizon. Our calculation accounting for the third-order
relativistic effects is in essence equivalent to 
the linear-order calculation in \cite{DEDOGR15}, and it shares the physical
basis with those in \cite{PASCZA13,DAPASC15,DAPASC15a}, where the second-order
calculations are performed in the 
conformal Fermi coordinates.

Similar operations can be performed to compute $\delta_{2,s}(t,{\bm x})$. 
The quadratic terms in Eq.~(\ref{eq:fullsol12}) 
absorb the cubic terms in proportion to~$\delta_1^{(2)}$, leaving
the standard Newtonian solution~$\delta_2^{(2)}$. The remaining cubic
terms correspond to the relativistic correction computed in \cite{JEGOET11},
vanishing $\propto k^2$ in the large-scale limit, after the coordinate 
rescaling removes the long-mode contribution to a constant. 
The local observer in the proper-time hypersurface would, therefore,
 feel the nonlinear gravitational effect with 
$\tilde\Delta\tilde\zeta$. 
This proves that the nonvanishing correlations of long and 
short 
wavelength
modes, a signature of local-type non-Gaussianity, must originate from 
non-gravitational forces.

Our third-order relativistic solutions in 
Eqs.~(\ref{eq:fullsol11})$-$(\ref{eq:fullsol23}) provide useful tools to 
analyze higher-order statistics in the proper-time hypersurface
such as the bispectrum and the trispectrum,
accounting for the nonlinear relativistic effects set up by the comoving-gauge
curvature perturbation
at the initial epoch. It is only when the observable
quantities are computed that the coordinate rescaling is naturally 
performed, as is the case in the computation of the single-field 
consistency relation.

\acknowledgments
We acknowledge useful discussions with Mehrdad Mirbabayi and David Wands.
We are especially grateful to Matias Zaldarriaga for the critical comments
on the manuscript and Jai-chan Hwang for sharing his unpublished 
results, to which we compare our results.
J.Y. is supported by the Swiss National Science Foundation. J.G. 
is supported by the Independent Junior Research Fellowship and 
by a Starting Grant through the Basic Science Research Program 
of the National Research Foundation of Korea (2013R1A1A1006701).

\vfill

\bibliography{ms.bbl}

\end{document}